\documentclass[11pt,british,conference,twocolumn]{IEEEtran}
\usepackage[T1]{fontenc}
\usepackage[latin9]{inputenc}
\usepackage{graphicx}

\makeatletter
\newenvironment{lyxcode}
{\par\begin{list}{}{
\setlength{\rightmargin}{\leftmargin}
\setlength{\listparindent}{0pt}% needed for AMS classes
\raggedright
\setlength{\itemsep}{0pt}
\setlength{\parsep}{0pt}
\normalfont\ttfamily}%
 \item[]}
{\end{list}}

\newcommand{\realign}{$\!\!\!\!\!\!\!\!\!\!$}

\makeatother

\usepackage{babel}
\begin{document}

\title{Speex: A Free Codec For Free Speech}

\author{Jean-Marc Valin\\
CSIRO ICT Centre, Cnr Vimiera \& Pembroke Roads, Marsfield NSW 2122,
Australia\\
Xiph.Org Foundation\\
\texttt{jean-marc.valin@csiro.au}}
\maketitle
\begin{abstract}
The Speex project has been started in 2002 to address the need for
a free, open-source speech codec. Speex is based on the Code Excited
Linear Prediction (CELP) algorithm and, unlike the previously existing
Vorbis codec, is optimised for transmitting speech for low latency
communication over an unreliable packet network. This paper presents
an overview of Speex, the technology involved in it and how it can
be used in applications. The most recent developments in Speex, such
as the fixed-point port, acoustic echo cancellation and noise suppression
are also addressed.\end{abstract}


\section{Introduction}

When I started implementing Speex in 2002, there was no good free
codec that could be used for speech communication. The situation was
greatly harming the development of voice over IP (VoIP) on Linux and
other free operating systems. Unlike the Vorbis\footnote{http://www.vorbis.com/}
codec that already existed at that time, the Speex codec is optimised
for speech and is designed for low latency communication over an unreliable
packet network.

Speex is based on the popular Code Excited Linear Prediction (CELP)
algorithm but because of patent restrictions, some techniques like
the use of algebraic codes (ACELP) could not be used. Nevertheless,
Speex is able to achieve a quality comparable to proprietary codecs
at the same bitrate, while supporting features not found in other
speech codecs. These features include variable bitrate, which makes
better use of the bits when encoding to a file, and embedded coding,
which provides an easy way to interface wideband (16 kHz) channels
with legacy narrowband (8 kHz) telephony equipment. 

Speex is now evolving into a complete toolkit for voice over IP (VoIP)
development, including algorithms for noise cancellation, acoustic
echo cancellation and adaptive jitter buffering. This allows a developer
without any signal processing knowledge to implement a VoIP client.
In the mean time, Speex is being ported to architectures without a
floating-point unit, allowing Speex to be used in embedded devices
equipped with a fixed-point CPU or DSP.

The paper is organised as follows. Section \ref{sec:Overview} presents
overview of Speex. Then, an explanation of the underlying coding algorithm
(CELP) is provided in Section \ref{sec:Speex-and-CELP}. Section \ref{sec:Implementing-Speex-Support}
provides information and tips about using the Speex library in applications.
Finally, Section \ref{sec:Recent-Developments} discusses recent developments
in Speex and Section \ref{sec:Conclusion} concludes this paper.

\section{Overview\label{sec:Overview}}

The Speex project was stated in February 2002 because there was a
need for a speech codec that was open-source and free from software
patents. These are essential conditions for being used by any open-source
software. At that time, the only options for compressing speech were
the G.711 codec (also known as $\mu$-law and A-law) and the open-source
implementation of the aging GSM-FR codec\footnote{http://kbs.cs.tu-berlin.de/\textasciitilde{}jutta/toast.html},
although the patent status of the latter is unclear. The Vorbis codec
was also available for general audio compression, but it was not really
suitable for speech. 

Unlike many other speech codecs, Speex is not targeted at cell phones
but rather at voice over IP (VoIP) and file-based compression. Because
any codec suitable for VoIP is also suitable for file-based compression,
only the VoIP requirements are taken into account. Version 1.0 was
released in March 2003 and the bitstream has been frozen since that
time, which means that all versions since then are compatible with
version 1.0.

\subsection{Design Decisions}

The use of Speex for VoIP imposes the following requirements:
\begin{itemize}
\item The frame size and algorithmic delay must be small
\item Both the encoder and decoder must run in real-time with limited resources
\item The effect of lost packets during transmission must be minimised
\item The codec must support both narrowband and wideband
\item Multiple bit-rates and quality settings must be supported to take
into account different connection speeds
\item Good compression must be achieved while avoiding known speech coding
patents
\end{itemize}
The CELP algorithm \cite{Schroeder1985} was chosen because of its
proven track record from low bitrates (e.g. DoD CELP at 4.8 kbps)
to high bitrates (e.g. G.728 LD-CELP at 16 kbps) and because the patents
on the base algorithm have now expired. Unfortunately, newer variants
such as the use of arithmetic codebooks in ACELP could not be used
in Speex because of patent issues. It was decided to encode in chucks
of 20 ms (frame size) and only use 10 ms of additional buffering (look-ahead). 

Because Speex is designed for VoIP instead of cell phones (like many
other speech codecs) it must be robust to lost packets, but not to
corrupted ones -- UDP/RTP packets either arrive unaltered or they
don't arrive at all. For this reason, we have chosen to restrict the
amount of inter-frame dependency to pitch prediction only, without
incurring the overhead of coding frames independently as done in the
iLBC \cite{Andersen2002} codec.

As much as we would like it to be true, Speex is not the most technologically
advanced speech codec available and it was never meant to be. The
original goal was to achieve the same quality as state-of-the-art
codecs with comparable or slightly higher bit-rate and this goal has
been achieved.

Speex also has the following features, some of which are not found
in other speech codecs:
\begin{itemize}
\item Integration of narrowband and wideband using an embedded bit-stream
\item Wide range of bit-rates available (from 2 kbps to 44 kbps)
\item Dynamic bit-rate switching and Variable Bit-Rate (VBR)
\item Voice Activity Detection (VAD, integrated with VBR)
\item Variable complexity
\item Ultra-wideband mode at 32 kHz
\item Intensity stereo encoding option
\end{itemize}

\subsection{Legal Status}

One of the most frequently asked question about Speex is: ``But how
can you be sure it is not patented?''. The short answer is that we
don't know for sure. The longer answer is that first, we (Xiph.Org,
me) obviously did not patent Speex. During development, we were careful
not to use techniques known to be patented and we searched through
the USPTO database to see if anything we were doing was covered. 

That being said, we cannot offer any absolute warranty regarding patents.
No company does that, simply because of the mess the patent system
is. Even if you pay for a codec, you cannot have the guaranty that
nobody else owns a (previously unknown) patent on that codec. Even
if someone comes forward and claims to have a patent on a certain
piece of software, it takes years of legal procedures to sort out.
This has been happening recently with the JPEG standard, which has
always been thought to be free of patents. It is not yet clear what
will happen in this particular case. Sadly, this is how things are
when it comes to patents -- at least in the United States because
every country has different patent laws, which makes the matter even
more complicated.

In any program written that is longer than \textquotedbl{}Hello World\textquotedbl{}
(and even that!), there might be a patent issue, generally a trivial
patent (e.g. 1-click shopping) nobody thought about. There are probably
patents out there claiming rights on \textquotedbl{}using a computer
for doing useful things\textquotedbl{} or something like that. What
this means for Speex is that we did our best to avoid known patents
and minimise the risk. This is about the best that can be done when
writing software at the moment, regardless of whether the software
is Free or proprietary.

\section{Speex and CELP\label{sec:Speex-and-CELP}}

Speex is based on CELP, which stands for Code Excited Linear Prediction.
The CELP technique is based on three ideas:
\begin{enumerate}
\item Using a linear prediction (LP) model to model the vocal tract;
\item Using of an adaptive and a fixed codebook as the input (excitation)
of the LP model;
\item Performing a search in closed-loop in a ``perceptually weighted domain''.
\end{enumerate}
The description of the Speex encoder and decoder in this section has
been greatly simplified for clarity purposes. Nonetheless, speech
processing knowledge is required to understand some parts.

\subsection{Source-Filter Speech Model}

Before going into Speex and CELP in details, let us introduce the
source-filter model of speech. This model assumes that the vocal cords
are the source of spectrally flat sound (the ``excitation signal''),
and that the vocal tract acts as a filter to spectrally shape the
various sounds of speech. While still an approximation, the model
is widely used in speech coding because of its simplicity. Its use
is also the reason why most speech codecs (Speex included) perform
badly on music signals.

The different phonemes can be distinguished by their excitation (source)
and spectral shape (filter). Voiced sounds (e.g. vowels) have an excitation
signal that is periodic and that can be approximated by an impulse
train in the time domain or by regularly-spaced harmonics in the frequency
domain. On the other hand, fricatives (such as the ``s'', ``sh''
and ``f'' sounds) have an excitation signal that is similar to white
Gaussian noise. So called ``voice fricatives'' (such as ``z''
and ``v'') have excitation signal composed of an harmonic part and
a noisy part.

\subsection{Speex Decoder}

\begin{figure}
\begin{center}\includegraphics[width=1\columnwidth,keepaspectratio]{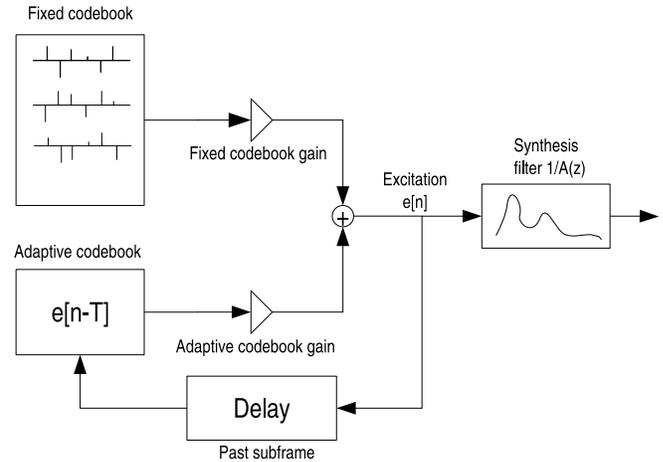}\end{center}

\caption{A generic CELP decoder.\label{cap:A-generic-CELP-decoder}}
\end{figure}

Before exploring the complex encoding process of Speex we introduce
the Speex decoder here. For the sake of simplicity only the narrowband
decoder is presented. Figure \ref{cap:A-generic-CELP-decoder} describes
a generic CELP decoder. The excitation is produced by summing the
contributions from an adaptive (\emph{aka} pitch) codebook and a fixed
(\emph{aka} innovation) codebook:
\begin{equation}
e[n]=e_{a}[n]+e_{f}[n]\label{eq:adaptive+fixed}
\end{equation}
where $e_{a}[n]$ is the adaptive codebook contribution and $e_{f}[n]$
is the fixed codebook contribution. 

The filter that shapes the excitation has an all-pole (infinite impulse
response) model of the form $1/A(z)$, where $A(z)$ is called the
prediction filter and is obtained using linear prediction (Levinson-Durbin
algorithm). An all-pole filter is used because it is a good representation
of the human vocal tract and because it is easy to compute.

In Speex, frames are 20 ms long, which is 160 samples for narrowband.
Furthermore, each frame is divided into 4 sub-frames of 40 samples
that are encoded sequentially. In most modes, only the synthesis filter
and the global excitation gain (see below) is encoded on a frame basis,
the other parameters are encoded on a sub-frame basis.

Speex includes three main differences compared to most recent CELP
codecs. First, while most recent CELP codecs make use of fractional
pitch estimation \cite{Kroon1990} with a single gain, Speex uses
an integer to encode the pitch period, but uses a 3-tap predictor
\cite{Chen1995} (3 gains). The adaptive codebook contribution can
thus be expressed as:
\begin{equation}
e_{a}[n]=g_{0}e[n-T-1]+g_{1}e[n-T]+g_{2}e[n-T+1]\label{eq:adaptive-3tap}
\end{equation}
where $g_{0}$, $g_{1}$ and $g_{2}$ are the pitch gains and are
jointly quantised (VQ). 

Many current CELP codecs use moving average (MA) prediction to encode
the fixed codebook gain. This provides slightly better coding at the
expense of introducing some extra dependency with past frames. Speex
encodes the fixed codebook gain as a global excitation gain for the
frame as well as a set of sub-frame gain corrections.

Speex uses sub-vector quantisation of the innovation (fixed codebook)
signal. Unfortunately, this is not a very efficient quantisation method,
but it was the best that could be used given patent restrictions.
Each sub-frame is divided into sub-vectors of length ranging between
5 and 20 samples. Each sub-vector is chosen from a bitrate-dependent
codebook and all sub-vectors are then concatenated. As an example,
the 3.95 kbps mode uses a sub-vector size of 20 samples with 32 entries
in the codebook (5 bits). This means that the innovation is encoded
with 10 bits per sub-frame, or 2000 bps. On the other hand, the 18.2
kbps mode uses a sub-vector size of 5 samples with 256 entries in
the codebook (8 bits), so the innovation uses 64 bits per sub-frame,
or 12800 bps.

\subsection{Speex Encoder}

The main principle behind CELP is called Analysis-by-Synthesis (AbS)
and means that the encoding (analysis) is performed by perceptually
optimising the decoded (synthesis) signal in a closed loop. In theory,
the best CELP stream would be produced by trying all possible bit
combinations and selecting the one that produces the best-sounding
decoded signal. This is obviously not possible in practice for two
reasons: the required complexity is beyond any currently available
hardware and the ``best sounding'' selection criterion implies a
human listener. 

In order to achieve real-time encoding using limited computing resources,
the CELP optimisation is broken down into smaller, more manageable,
sequential searches using a simple perceptual weighting function.
In the case of Speex, the optimisation is performed in four steps
\begin{enumerate}
\item Linear prediction analysis is used to determine the synthesis filter,
which is converted to Line Spectral Pair (LSP) coefficients and vector-quantised.
\item The adaptive codebook entry and gain are jointly searched for the
best pitch-gain combination using AbS.
\item The fixed codebook gain is determined in an ``open-loop'' manner
only the energy of the excitation signal.
\item The fixed codebook is searched for the best entry using AbS.
\end{enumerate}
Optimisation for steps 2) and 4) are performed in the so called ``perceptually
weighted domain'', meaning that we try to minimise the perceptual
difference with the original (as opposed to simply maximising the
signal-to-noise ratio). In order to do that, the following weighting
filter is applied on the input signal:
\begin{equation}
W(z)=\frac{A(z/\gamma_{1})}{A(z/\gamma_{2})}\label{eq:weighting-filter}
\end{equation}
where $A(z)$ is the linear prediction filter and $\gamma_{1}$ and
$\gamma_{2}$ control the shape of the filter by moving the poles
and zeros toward the centre of the \emph{z}-transform unit circle
(Speex uses $\gamma_{1}=0.9$ and $\gamma_{2}=0.6$). The filter defined
in \ref{eq:weighting-filter} is in fact a very, very rough approximation
for what is known as the ``masking curve'' in audio codecs such
as Vorbis. The overall effect of the weighting filter $W(z)$ is that
the encoder is allowed to introduce more noise at frequencies where
the power level is high and less noise at frequencies where the power
level is low.

\subsection{Wideband CELP (aka SB-CELP)}

The wideband mode in Speex differs from most wideband speech codecs
as it separates the signal into two sub-bands (SB-CELP). The lower
band is encoded using the narrowband encode. This has the advantage
of making it easy to interoperate with ``legacy'' systems (e.g.
PSTN) operating in narrowband. The high-band is encoded using a CELP
encoder similar to the narrowband encoder with the difference that
only a fixed codebook is used, with no adaptive codebook. The high-band
information is encoded after then narrowband (low-band) data, so simply
discarding the last part of a wideband frame produces a valid narrowband
frame. Alternatively, a wideband decoder is able to identify that
the high-band information is missing and thus is able to decode a
narrowband frame. 

One particularity of the wideband mode made possible by the SB-CELP
scheme is that for low-bitrate, only the spectral shape (envelope)
of the high-band is encoded. The high-band excitation is obtained
simply by ``folding'' the narrowband excitation. The technique is
similar to the low bit-rate encoding of the high frequencies described
in \cite{Valin2000}.

\section{Implementing Speex Support\label{sec:Implementing-Speex-Support}}

Unlike most speech codecs, Speex supports several sampling rates and
several bit-rates for each sampling rate. The thing one must determine
before choosing a codec (or in this case, which mode of the codec)
is:
\begin{itemize}
\item What is the bandwidth available?
\item What is the quality requirement?
\item What is the latency requirement?
\end{itemize}
When designing a Voice over IP application, the first aspect one must
consider is the fact that the combination of IP, UDP and RTP incurs
an overhead of 40 bytes per packet transmitted. When transmitting
one packet every 20 ms as is commonly done, the overhead is equal
to 16 kbps! Considering that, it is obvious that in most situations,
there is little incentive to use very low bit-rate.

\subsection{Choosing the Right Mode}

The first choice one must make when using Speex is the sampling rate
used. For new applications, it is recommended to use wideband (16
kHz sampling rate) audio whenever possible. Wideband speech sounds
much clearer than narrowband speech and is also easier to understand.
In many case, for example when implementing a VoIP application, it
is a good idea to also support narrowband -- either directly or through
the wideband encoder/decoder. In this context, narrowband is useful
for both compatibility reasons (with clients that may not support
wideband) and because at very low bit-rates (<10 kbps), it sometimes
provides better quality than wideband. 

Once the sampling rate is chosen, one must decide on the bitrate.
It is generally recommended to use constant bitrate (CBR) in voice
over IP applications because of the fixed channel (especially if using
a modem) capacity. On the other hand, VBR is usually more efficient
for file compression, because bits can be allocated where needed.
When calculating for a target bit-rate, it is important to also consider
the overhead. The VoIP overhead when using RTP and one frame per packet
is 16 kbps, so when using a 56 kbps modem, this leaves very little
bandwidth for the codec itself. For file encoding in Ogg, the overhead
is approximately 1 byte per frame, so it is usually not problematic.
For very low bit-rates, the \emph{--nframes} option to \emph{speexenc}
can be useful as it packs multiple frames in an Ogg packet.

\subsection{Libspeex API Overview}

The reference implementation of Speex uses a C API. Only an overview
of the codec API is provided here. When using Speex, one must first
create an instance of an encoder and/or a decoder. If more than one
channel is used, then an instance is required for each channel (Speex
is not stateless). 

In order to encode speech using Speex, the following header must be
included:
\begin{lyxcode}
{\footnotesize{}{\realign}\#include~<speex/speex.h>}{\footnotesize \par}
\end{lyxcode}
Then, a Speex bit-packing struct and a encoder state are necessary:
\begin{lyxcode}
{\footnotesize{}{\realign}SpeexBits~bits;}{\footnotesize \par}

{\footnotesize{}{\realign}void~{*}enc;}{\footnotesize \par}
\end{lyxcode}
These are initialized by:
\begin{lyxcode}
{\footnotesize{}{\realign}speex\_bits\_init(\&bits);}{\footnotesize \par}

{\footnotesize{}{\realign}enc~=~speex\_encoder\_init(\&speex\_nb\_mode);}{\footnotesize \par}
\end{lyxcode}
For wideband coding, \texttt{\footnotesize{}speex\_nb\_mode} will
be replaced by \texttt{\footnotesize{}speex\_wb\_mode}. In most cases,
you will need to know the frame size used by the mode you are using.
The value can be obtained in the \texttt{\footnotesize{}frame\_size}
variable using:
\begin{lyxcode}
{\footnotesize{}{\realign}speex\_encoder\_ctl(enc,}{\footnotesize \par}

{\footnotesize{}SPEEX\_GET\_FRAME\_SIZE,\&frame\_size);}{\footnotesize \par}
\end{lyxcode}
In practice, \texttt{\footnotesize{}frame\_size} will correspond to
20 ms when using 8, 16, or 32 kHz sampling rate.

Once the initialization is complete, frames can be encoded using:
\begin{lyxcode}
{\footnotesize{}{\realign}speex\_bits\_reset(\&bits);}{\footnotesize \par}

{\footnotesize{}{\realign}speex\_encode\_int(enc\_state,~input\_frame,~}{\footnotesize \par}

{\footnotesize{}\&bits);}{\footnotesize \par}

{\footnotesize{}{\realign}nbBytes~=~speex\_bits\_write(\&bits,~byte\_ptr,~}{\footnotesize \par}

{\footnotesize{}MAX\_NB\_BYTES);}{\footnotesize \par}
\end{lyxcode}
where \texttt{\footnotesize{}input\_frame} is a \emph{(int16\_t {*})}
pointing to the beginning of a speech frame, \texttt{\footnotesize{}byte\_ptr}
is a \emph{(char {*})} where the encoded frame will be written, \texttt{\footnotesize{}MAX\_NB\_BYTES}
is the maximum number of bytes that can be written to \texttt{\footnotesize{}byte\_ptr}
without causing an overflow and \texttt{\footnotesize{}nbBytes} is
the number of bytes actually written to \texttt{\footnotesize{}byte\_ptr}
(the encoded size in bytes). Before calling \texttt{\footnotesize{}speex\_bits\_write},
it is possible to find the number of bytes that need to be written
by calling \texttt{\footnotesize{}speex\_bits\_nbytes(\&bits)}, which
returns a number of bytes.

After you're done with the encoding, all resources can be freed using:
\begin{lyxcode}
{\footnotesize{}{\realign}speex\_bits\_destroy(\&bits);}{\footnotesize \par}

{\footnotesize{}{\realign}speex\_encoder\_destroy(enc\_state);}{\footnotesize \par}
\end{lyxcode}
The decoder API can be used in a similar manner. For more details,
refer the Speex manual\footnote{http://www.speex.org/docs.html}.

\subsection{Common Layer 8 Problems}

There are problems that frequently arise when implementing support
for Speex. While some of them are due to programming errors, most
are by sending or receiving the wrong signals. The most important
aspect for the input signal is the amplitude. Speex can encode any
signal in the $[-32768,32767]$ range. However, it is best to make
use of the full dynamic range without being too close to saturation
because that could cause the decoded speech to saturate even if the
encoded speech does not. In practice, it is best to scale the input
signal so the maximum amplitude is between 5000 and 20000.

Another important characteristics of the signals is the frequency
content. DC offsets should be avoided at all cost as it causes the
Speex encoder to produce bad sounding files (DC is never needed for
audio). A DC offsets is sometimes caused by a cheap soundcard and
means that the signal is not centred around the zero (zero-mean).
In the most simple case, a DC offset can easily be removed using a
notch filter of the form:
\begin{equation}
N(z)=\frac{1-z^{-1}}{1-\alpha z^{-1}}\label{eq:notch_filter_z}
\end{equation}
where a value of $\alpha=0.98$ is usually appropriate for both narrowband
and wideband. The filter described in (\ref{eq:notch_filter_z}) can
be implemented as:
\begin{lyxcode}
{\footnotesize{}{\realign}}\#define~ALPHA~.98f

{\footnotesize{}{\realign}}static~float~mem=0;

{\footnotesize{}{\realign}}for~(i=0;i<frame\_size;i++)

{\footnotesize{}{\realign}}\{

{\footnotesize{}{\realign}}$\:\:\:$mem~=~ALPHA{*}mem~+~(1-ALPHA){*}input{[}i{]};

{\footnotesize{}{\realign}}$\:\:\:$output{[}i{]}~=~output{[}i{]}~-~mem;

{\footnotesize{}{\realign}}\}
\end{lyxcode}
While the effect is less obvious, sometimes excessive amount of low
frequencies can cause sub-optimal encoding with Speex. As a general
rule, narrowband audio should be high-pass filtered at 300 Hz and
wideband audio should be filtered at 50 Hz. It is one of the planned
features for Speex to do the filtering automatically, this has not
been done yet (as of November 2005).

\section{Recent Developments\label{sec:Recent-Developments}}

Improvements to Speex have continued since the Speex bitstream was
frozen with the release of version 1.0. These improvements are included
in the 1.1.x unstable branch, which will lead to version 1.2. These
are improvements to the actual implementation as well as extra features
outside of the codec itself. It should be emphasised that the bitstream
of version 1.2 will be exactly the same as for version 1.0.

\subsection{Fixed-Point Port}

Since November 2003 (version 1.1.1), work has been going on to convert
Speex to fixed-point. This effort has been pushing Speex into embedded
devices.

While the port is not yet complete (as of November 2005), a large
part of the code has now been converted. Because the fixed-point code
is implemented using arithmetic operator abstraction (macros), using
Speex on a fixed-point architecture is as simple as configuring with
the --enable-fixed-point option or defining the FIXED\_POINT macro.
As of November 2005, the status of the different components of Speex
regarding fixed-point are as follows:
\begin{itemize}
\item Narrowband: Completely converted for all constant bit-rates ranging
from 3.95 kbps to 18.2 kbps (inclusively). Other bit-rate and narrowband
options have only a few float operations left (float emulation is
realistic)
\item Wideband: Mostly converted to fixed-point, but some float operations
are left (float emulation is realistic on a fast chip).
\item Preprocessor: Not converted to fixed-point yet.
\item Echo canceller: Fixed-point port planned in the near future.
\end{itemize}
Using the fixed-point port, Speex is known to run on the following
fixed-point architectures:
\begin{itemize}
\item Blackfin ($\mu$Clinux)

Speex now includes architecture-dependent optimisations for the Analog
Devices Inc. Blackfin DSP. It is possible to develop with Speex on
Blackfin using $\mu$Clinux\footnote{http://blackfin.uclinux.org/},
GCC and Linphone\footnote{http://www.linphone.org/} on the GPL-licensed
STAMP board\footnote{http://blackfin.uclinux.org/projects/stamp/}.

\item ARM (Linux, Symbian)

Speex includes architecture-dependent optimisations for the ARM v4
and ARM v5E architectures. The GCC compiler can be used to compile
Speex on ARM.

\item TI C55x, C54x, C6x.

Speex has been adapted to the ``non-conventional'' architecture
of these TI DSP families. There are unfortunately no architecture-dependent
optimisations yet and (to the author's knowledge) no Free compiler
available.

\end{itemize}
Other fixed-point architectures may also be able to run Speex simply
by enabling fixed-point compilation.

\subsection{Preprocessor}

A preprocessor module was introduced early in the 1.1.x branch. It
is still under development, but many developers have already used
it successfully. The preprocessor is designed to be used on the audio
\emph{before} running the encoder and provides three functionalities:
\begin{itemize}
\item noise suppression (denoiser)
\item automatic gain control (AGC)
\item voice activity detection (VAD)
\end{itemize}
The denoiser can be used to reduce the amount of background noise
present in the input signal. This provides higher quality speech whether
or not the denoised signal is encoded with Speex (or at all). However,
when using the denoised signal with the codec, there is an additional
benefit. Speech codecs in general (Speex included) tend to perform
poorly on noisy input and tend to amplify the noise. The denoiser
greatly reduces this effect.

Automatic gain control (AGC) is a feature that deals with the fact
that the recording volume may vary by a large amount between different
hardware and software setups. The AGC provides a way to adjust a signal
to a reference volume. This is useful for voice over IP because it
removes the need for manual adjustment of the microphone gain. A secondary
advantage is that by setting the microphone gain to a conservative
(low) level, it is easier to avoid clipping.

The voice activity detector (VAD) provided by the preprocessor is
more advanced than the one directly provided in the codec. Unfortunately,
it cannot directly be used by the encoder yet.

\subsection{Acoustic Echo Cancellation}

The Speex library now includes an echo cancellation algorithm suitable
for Acoustic Echo Cancellation (AEC). AEC is especially useful in
applications (VoIP or others) where one records from a microphone
while speakers are being used. Speex uses an implementation of the
multidelay block frequency (MDF) algorithm \cite{Soo1990}. Again,
the algorithm is still under development, but has already been successfully
used in some applications.

Echo cancellation algorithms are generally very dependent on the quality
and characteristics of the processed signals. The Speex AEC is no
exception and there are several things that may prevent the echo canceller
from working properly, some of which are non-obvious to people with
no signal processing background. While it is always possible to encounter
a bug in the Speex AEC, there are many other options that should be
considered before. 

First, using a different soundcard to do the capture and playback
will not work, regardless of what some people think. The only exception
to that is if the two cards can be made to have their sampling clock
``locked'' on the same source. However, this feature is only supported
on professional soundcards.

The delay between the record and playback signals must be minimal.
Any signal played has to ``appear'' on the playback (far end) signal
slightly before the echo canceller ``sees'' it in the near end signal.
However, excessive delay means that part of the filter length is wasted.
In the worst-case situation, the delay is such that it is longer than
the filter length, in which case, no echo can be cancelled.

When it comes to echo tail length (filter length), longer is not necessarily
better. The longer the tail length, the longer it takes for the filter
to adapt. Of course, a tail length that is too short will not cancel
enough echo, but the most common problem seen is that people set a
very long tail length and then wonder why no echo is being cancelled
(or why adaptation is very slow).

Non-linear distortion cannot (by definition) be modelled by the linear
adaptive filter used in the echo canceller and thus cannot be cancelled.
It is thus best to use good audio hardware and avoid saturation/clipping
at all cost. Also, the AEC should be the first algorithm to process
the microphone signal and the last algorithm to see the speaker (far
end) signal. A common mistake when implementing echo cancellation
is to use noise suppression or automatic gain control before the AEC.
Any processing, especially if it is non-linear, should occur only
after the echo cancellation algorithm.

\section{Conclusion\label{sec:Conclusion}}

In this paper, the origin and design goals of the Speex speech codec
were presented. Also, a description of the CELP algorithm and its
implementation in Speex was given. Some guidelines for programming
with libspeex and choosing the right encoding options were provided
in order to help developers make better use of Speex in applications. 

Even though the bitstream itself has been frozen for more than two
years, work on Speex is continuing. Recent improvements focus on bringing
support for fixed-point CPU and DSP architectures as well as providing
additional functionality, such as echo cancellation and noise suppression,
that are useful for VoIP applications.

\section*{Resources}

Speex: http://www.speex.org 

Speex mailing list: speex-dev@xiph.org 

Xiph.Org Foundation: http://www.xiph.org

\section*{Application support}

Speex is supported in the following applications\footnote{Only open-source software is listed here, in no particular order.
This is not a complete list.}:
\begin{itemize}
\item OpenH323 (http://www.openh323.org/): Free implementation of the H323
protocol
\item Linphone (http://www.linphone.org/): SIP-based VoIP application, one
of the first application to support Speex
\item Gnomemeeting (http://www.gnomemeeting.org/): VoIP application based
on OpenH323 (with SIP support coming)
\item Sweep (http://www.metadecks.org/software/sweep/): Sound editor
\item Asterisk (http://www.asteriskpbx.org/): PBX Software
\item Annodex (http://www.annodex.net/): Open standard for annotating media
\item Ogg DirectShow (http://www.illiminable.com/ogg/): Windows DirectShow
filters for Speex, Vorbis, FLAC and Theora
\item JSpeex (http://jspeex.sourceforge.net/): Java port of Speex
\item Libfishsound (http://www.annodex.net/software/libfishsound/): Simple
library for encoding and decoding Ogg audio files
\item GStreamer (http://gstreamer.freedesktop.org/): Multimedia framework
library
\item MPlayer (http://www.mplayerhq.hu/): Media player
\item Xine (http://xinehq.de/): Media player
\item VLC (http://http://www.videolan.org/vlc/): Media player and streaming
server
\item SIP Express Media Server (http://www.iptel.org/sems/): Extensible
media server for voice services
\end{itemize}
\bibliographystyle{ieeetr}
\bibliography{speex}

\end{document}